\documentclass[10pt, a4paper, onecolumn, conference, compsocconf]{IEEEtran}
\ifCLASSINFOpdf
\else
\fi
\usepackage{times}                               
\usepackage{graphicx}
\usepackage[cmex10]{amsmath}
\usepackage{amsthm,amssymb}

\newcommand{\procbegin}{\vspace*{3mm}\hrule\vspace{2mm}\noindent}
\newcommand{\procend}{\vspace*{2mm}\hrule\vspace{3mm}}

\newcommand{\hstt}{\\\hspace*{4mm}}

\newcommand{\hstttt}{\\\hspace*{8mm}}

\newcommand{\hstttttt}{\\\hspace*{12mm}}

\newcommand{\hstttttttt}{\\\hspace*{16mm}}

\newenvironment{list1}{\begin{list}{$\bullet$}
{\topsep 0 pt \parsep 0 pt \partopsep 0 pt \itemsep 0
pt}}{\end{list}}

\newcommand{\mcla}{\mathcal{A}}
\newcommand{\emvadd}{\mbox{\em vertex\_add}}

\newcommand{\emcadd}{\mbox{\em clock\_add}}

\newcommand{\emvaradd}{\mbox{\em var\_add}}
\newcommand{\emjadd}{\mbox{\em jump\_add}}
\newcommand{\emjedt}{\mbox{\em jump\_edit}}

\newtheorem{defi}{Definition}

\newtheorem{prop}{Proposition}

\newtheorem{obser}{Observation}

\begin{document}
%
\title{Modeling and Verification for Timing Satisfaction of Fault-Tolerant Systems with Finiteness}


\author{
\IEEEauthorblockN{Chih-Hong Cheng\authorrefmark{1},
                    Christian Buckl\authorrefmark{3},
                    Javier Esparza\authorrefmark{2}
                    Alois Knoll\authorrefmark{1}}
\IEEEauthorblockA{\authorrefmark{1}Unit 6: Robotics and Embedded
Systems, Department of Informatics, TU Munich, Germany}
\IEEEauthorblockA{\authorrefmark{1}Unit 11: Theoretical Computer Science, Department of Informatics, TU Munich, Germany}
\IEEEauthorblockA{\authorrefmark{2}Fortiss GmbH, Germany
\\Email:\{chengch,buckl,esparza,knoll\}@in.tum.de }}


%


\maketitle

\begin{abstract}
The increasing use of model-based tools enables further use of formal verification techniques in the context of distributed real-time systems. To avoid state explosion, it is necessary to construct verification models that focus on the aspects under consideration.

In this paper, we discuss how we construct a verification model for timing analysis in distributed real-time systems.
We (1) give observations concerning restrictions of timed automata to model these systems,
(2) formulate mathematical representations on how to perform model-to-model transformation to derive verification models from system models, and (3) propose some theoretical criteria how to reduce the model size. The latter is in particular important, as for the verification of complex systems, an efficient model reflecting the properties of the system under consideration is equally important to the verification algorithm itself.
Finally, we present an extension of the model-based development tool FTOS, designed to develop fault-tolerant systems, to demonstrate 
our approach.
\end{abstract}



%
\IEEEpeerreviewmaketitle
\section{Introduction}

The complexity of distributed real-time systems is growing rapidly; model-based development tools are used to accelerate the development process and increase the quality of the produced code. In addition, it is possible to integrate formal verification as analysis technique into these tools.


Currently, the standard verification process is achieved by first translating system models into verification models, followed by verifying relevant properties by verification engines using special algorithms. In the verification community, researchers focus on tighter theoretical complexity bounds or computationally faster algorithms to reduce the required time for verification. Nevertheless, if it comes to verification of complex systems, an efficient model reflecting the properties of the system under consideration becomes essential. By \emph{efficient model}, we refer to a model containing "just-enough" information of the system behavior regarding these properties. In fact, an inefficient modeling with irrelevant details can simply render the verification intractable.

Within this paper, we introduce an approach for the construction of such an efficient model for the verification of timing assumptions and constraints. The approach is presented in, but not restricted to, the context of FTOS~\cite{buckl:2008}, a model-based development tool for the design of fault-tolerant systems.

In our presentation, we first introduce FTOS, mention insights regarding differences in comparison to other development tools, and propose our two-phase verification methodology (sec.~\ref{sec.Motivating.Examples}).
Then based on FTOS and timed automata~\cite{alur:1994:tta}, we describe the model construction process, focusing on the aspects concerning expressiveness, modification, and efficiency.
\begin{itemize}
\item \textbf{(Expressiveness)} We give observations regarding restrictions of timed automata to construct models of real-time systems (sec.~\ref{sec.Network.Finite.Capacities},~\ref{sec.Task.Element.Finite.Precision},~\ref{sec.Dispatcher}); these observations are valid not only in the context of FTOS, but apply also for other systems.
\item \textbf{(Modification)} We formulate mathematical representations how to perform model modification to derive verification models from system models (sec.~\ref{sec.Job.Processing.Element}).
\item \textbf{(Efficiency)} With the understanding of (1) complexities of verification and (2) our problem structure, we propose some theoretical criteria regarding how to construct an efficient model, such that it is possible for existing model checkers to generate results within reasonable time (sec.~\ref{Sec.Invocation.Faults.Events}).
\end{itemize}

At last, we report our preliminary implementation (sec.~\ref{sec.Implementation}), mention related work (sec.~\ref{sec.Related.Work}), and conclude this paper (sec.~\ref{sec.Conclusion}).

\section{FTOS and Motivating Examples\label{sec.Motivating.Examples}}

\subsection{Introduction to FTOS}
FTOS is a model-based development tool for the development of fault-tolerant real-time systems, that alleviates designers' burden by offering code generation for non-functional aspects with high extensibility.

The conceptual modeling in FTOS uses multi-aspect techniques comprising four different perspectives:

\begin{list1}
    \item \textbf{Hardware Model}: The hardware model specifies the hardware used, including specifications of \emph{electronic control units} (ECUs) and the interconnecting \emph{network}.
    \item \textbf{Software Model}: The underlying model of computation in FTOS shares large similarities with that of Giotto \cite{Henzinger01giotto:a}, which is based on the concept of Logical Execution Times. A designer should specify \emph{tasks}, \emph{ports}, \emph{inputs}, \emph{outputs}, and \emph{jobs}.
    \item \textbf{Fault Model}: The fault model specifies the \emph{fault hypothesis} of the system, which includes the set of \emph{fault containment units} (FCUs) (possible faults concerning locations, types, durations), and the set of \emph{fault configurations}
                                (possible simultaneously activating FCUs). Examples for the fault hypothesis are:
                                \begin{enumerate}
\item A network link can have message lost (fault type: \verb"MsgLoss") with minimum interval between consecutive occurrences equal to $3$ milliseconds\footnote{This minimum interval is called \emph{least time between faults} (LTBF) in FTOS and is derived from the required probability of the system to withstand the fault.}.
\item A software task can produce errors (fault type: \verb"WrongResult") due to a fault within an associated sensor; once happened, it will not be corrected unless explicitly done by the user or the fault-tolerant mechanism. The minimum interval for the correct operation between two consecutive faults of the sensor is expected to be $500$ milliseconds\footnote{In (1) the message loss is transient, and in (2) computation errors caused by hardware faults are permanent.}.
\end{enumerate}
    \item \textbf{Fault-tolerance Model}: The fault-tolerance model specifies methods to detect errors and to repair and restore the system.
\end{list1}

During code generation, FTOS selects, adapts and combines pre-implemented code templates based on model features. A detailed description of FTOS can be found in \cite{buckl:2008}.

\begin{figure}
 \centering
 \includegraphics[width=0.45\columnwidth]{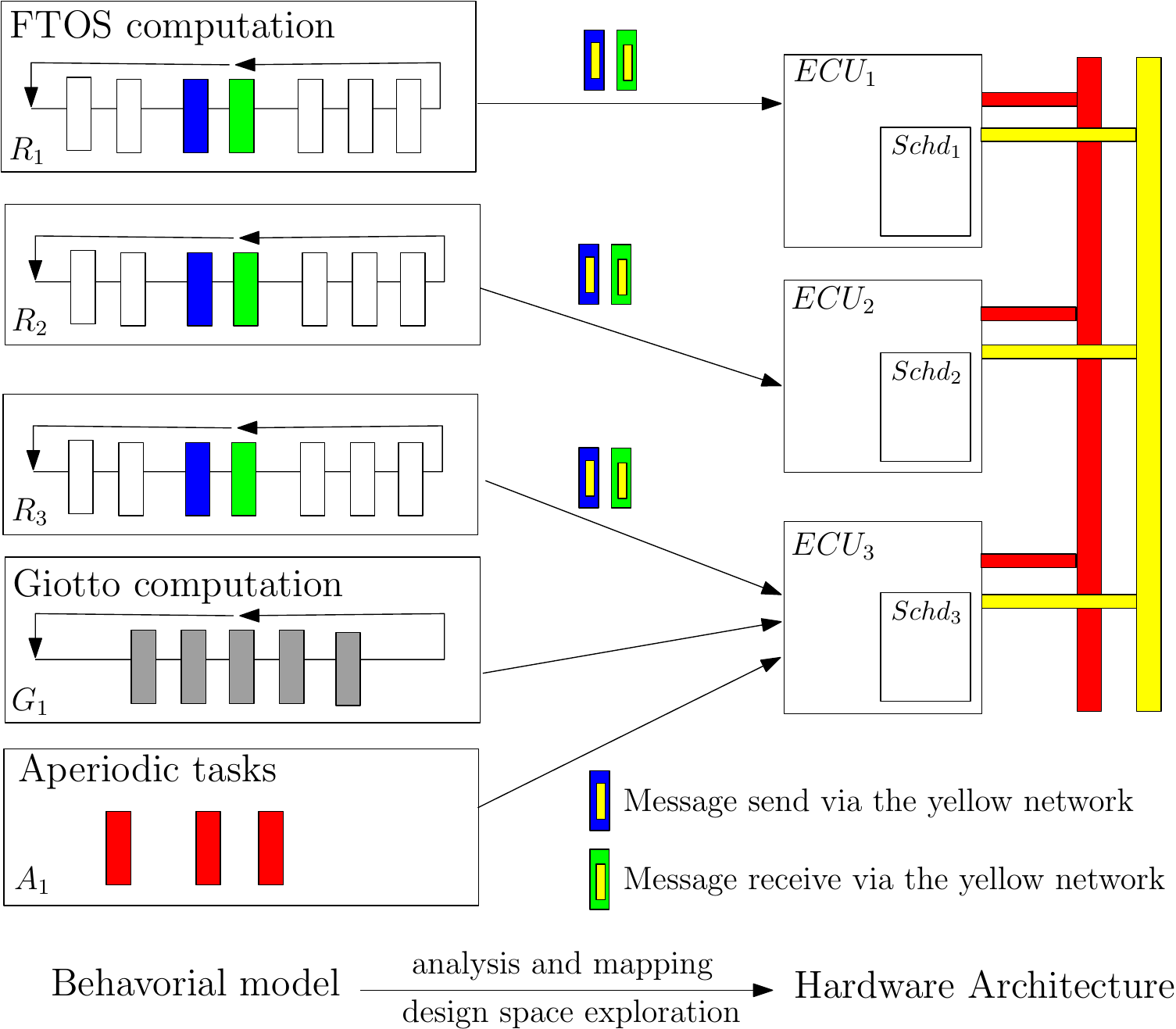}
 \caption{Behaviorial models and architectures.}
 \label{fig:Mapping}
\end{figure}

\subsection{General Settings and Examples}
The concepts presented in this paper do not only apply for FTOS, but also a range of other related projects, such as Giotto \cite{Henzinger01giotto:a} or event-driven tasks with fixed deadlines. Figure \ref{fig:Mapping} shows the different models of execution. An aperiodic or sporadic function is event driven; when such an event happens, a deadline is assigned to the task handling the event. Giotto functions are functions that interact synchronously at macro step level (logical level), while at micro step level the execution is asynchronous. For detailed description of Giotto and the concept of logical execution time, see \cite{Henzinger01giotto:a}. FTOS functions are extensions of Giotto functions. Intuitively they are equipped with fault-tolerance abilities such that the system can resist faults defined by the fault model. In fig.~\ref{fig:Mapping}, three redundant copies ($R_1$, $R_2$, $R_3$) are deployed on the three machines ($ECU_1$, $ECU_2$, $ECU_3$).

The figure also shows the necessity of a mapping the behavioral model (in FTOS: software model) to the architecture model (in FTOS: hardware model). Note that in general a design space exploration is needed for finding such a mapping. For details, we refer readers to articles regarding platform-based design \cite{sangiovannivincentelli:2001:pbd}. Since this mapping is specified in FTOS by the developer, our analysis can start from a given selection of hardware and software settings.

\subsection{Verification Goals\label{subsec.Verification.Goals}}
The main property of fault-tolerant systems that needs to be verified is the ability to withstand the assumed faults. The fault assumptions are summarized in the fault hypothesis (in FTOS: fault model) that defines faults regarding its location, effect, and frequency.

The verification of such systems is hindered by two aspects: deadline violations and non-determinism due to e.g. imperfect synchronization of redundant units.

\begin{figure}
 \centering
 \includegraphics[width=0.45\columnwidth]{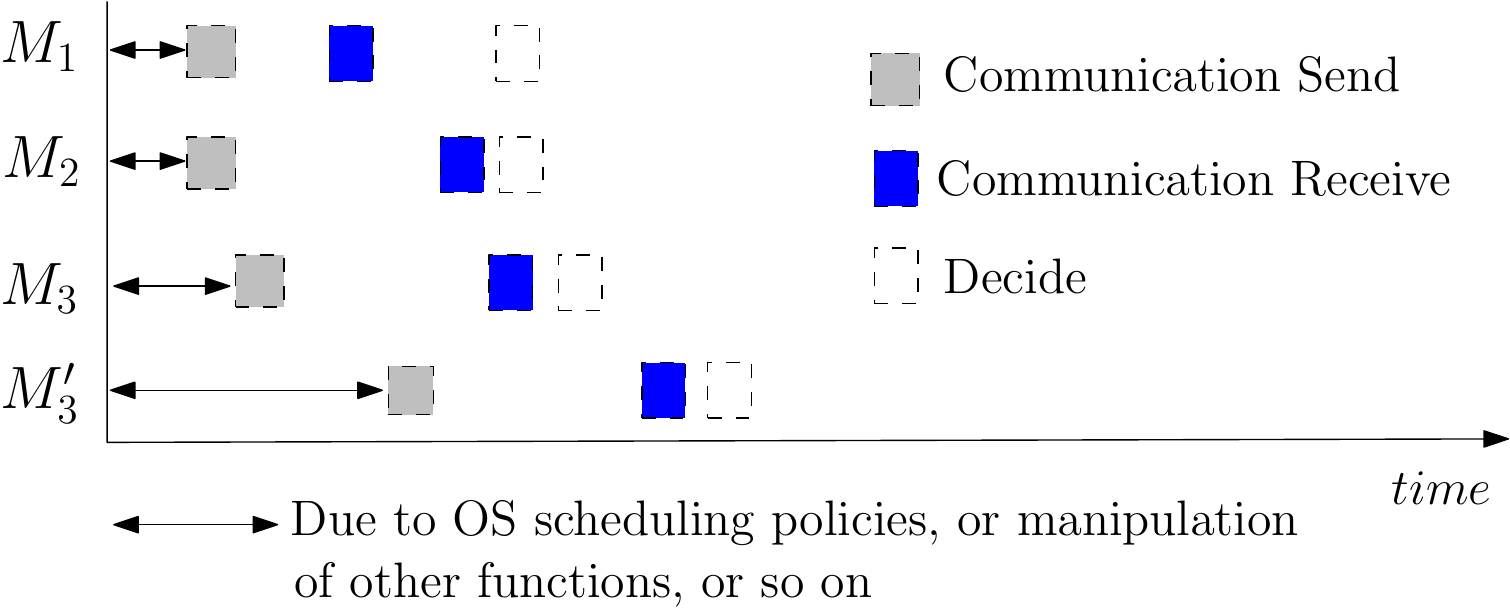}
 \caption{Internal nondeterminism due to scheduling differences.}
 \label{fig:Nondeterminism}
\end{figure}

\begin{enumerate}
\item In ordinary systems, correctness relies on the assumption that a scheduling never leads to deadline violations (without loss of generality, we assume that deadlines specified in our model are hard). Nevertheless, in fault-tolerant systems, the constraint can be loosened. Due to replication, a deadline violation of one unit might be tolerated. In fact, the violation of the deadline can be categorized as an occurrence of a fault defined in the fault model. This brings dramatic differences between fault-tolerant systems and ordinary systems, i.e., deadline violation is feasible or acceptable provided that there exists a fault-tolerance mechanism such that the effect of fault can be eliminated.
\item On the other hand, replication also introduces further difficulties. In ordinary Giotto systems, internal determinism is guaranteed, meaning that two deployments having the same relative ordering in the
micro step level will have the same behavior, irrelevent of the absolute timing. Unfortunately, internal determinism will not be maintained if
no constraints are added additionally on FTOS functions. Consider fig.~\ref{fig:Nondeterminism}, where $M_1$, $M_2$ and $M_3$ are three deployments.
The send action will broadcast messages to other machines regarding its liveness. Ideally, when no error happens, then each machine should conceive a consistent view of the system. However, when the scheduling of $M_3$ changes to that of $M_3'$, with zero time transmission, the result will be an inconsistent view at $M_1$ and $M_2$. This brings semantic incompatibility between
different deployments.
\end{enumerate}

To solve these problems, we thus propose the concept called \emph{deterministic assumption}~\cite{Cheng:2009:FTOSVerify}. Intuitively, the goal is to assume that the implementation of fault tolerance mechanisms will always provide a consistent view for all correct machines regardless of deadline violation  and scheduling issues. In practice, this will place constraints regarding the earliest and latest arrival time between messages sent, which need to be verified.

For above purpose, we adapt a two-phase verification process in our tool FTOS-Verify:
\begin{list1}
\item \textbf{(Phase 1: Verification on the platform independent layer)} We first assume that the deterministic assumption holds in all deployments. Based on this assumption, we construct a verification model. The model is an abstract machine (closed model) where injection of faults is regulated based on the fault model. The model offers precision by revealing detailed mechanisms of fault-tolerance. Our theoretical
      foundation enables us to construct a concise model with huge benefits\footnote{Our theorem states that we can construct a synchronous verification model (exponentially smaller reachable state space) provided that (1) the deterministic assumption holds and (2) the properties are local (in-machine) LTL properties without
      using temporal operator \textbf{X}. This makes formal verification of large systems practicable.}. For this phase, the mathematical formulation and the proof of theorems are stated in \cite{Cheng:2009:FTOSVerify}; it will not be the focus of this paper.

\item \textbf{(Phase 2: Validity checking of the behavior-architectural mappings)} In this phase, we have to focus on two aspects. First, we have to check whether the deterministic assumption holds in the platform. Second, we have to check if there exists possibilities where deadlines are violated, and the violation exceeds the constraint specified and regulated in the fault model. Note that since the correctness of the data and mechanisms are checked in the first phase, in the latter phase only protocol checking (timing) is needed. This will be the focus and the main contribution of the paper. For the analysis of the temporal behavior, we transform the models in FTOS to communicating timed automata (CTA). In the following sections, we will describe our observations, relevant parts of the construction process, and theoretical criteria for model efficiency. By using a generalized view, the results are applicable not only in the context of FTOS, but can be used for verifying temporal behavior for generic distributed real-time systems.
\end{list1}

\section{System Modeling and Observations\label{sec.System.Modeling.Observations}}
We use an extended format of communicating timed automata (CTA)~\cite{bdl:2004:uppaal} using variables of finite domain to express the features of the behavioral model. It is important to mention that this extended format does not change the expressiveness of CTA.

\begin{defi} A system of communicating timed automata is a tuple
$\mathcal{S}=\{\mathcal{A}_1, \ldots , \mathcal{A}_n\}$, where
$\mathcal{A}_i = (Q_i, V_i, C_i, Sync_i, q_i,$
 $Jump_i, Inv_i)$ is
an automaton with the following constraints.
\begin{list1}
\item $Q_i$ is a finite set of \emph{modes} (\emph{locations}).

\item $V_i$ is the set of finite-domain integer variables.

\item $C_i = \{c_{i_1},\ldots, c_{i_m}\}$ is the set of
\emph{clock variables}.

\item $Sync_i = \{s_{i_1}, \ldots, s_{i_n}\}$ is the set of
\emph{synchronizers}; each synchronizer $s$ is of the format $s
\in \{?,!\} \times \Sigma $ where elements in $\Sigma$ represent
synchronizer symbols. Conceptually, $"?"$ represents receiving,
and $"!"$ represents sending.

\item $q_i \in Q_i$ is the \emph{initial location} of the
automaton.

\item $Jump_i=Q_i\times Guards_i \times Sync_i \rightarrow Q_i \times Resets_i$
is the jump from mode to mode.
\begin{enumerate}
\item $Guards_i$ is the conjunction of inequalities
    of the form $c_{i_x} \sim k$ or $v_{j_x} \sim k'$, where $c_{i_x} \in
    C_i$, $v_{j_x} \in V_j, j=1\ldots n$, $k,k' \in \mathbb{N}$, and $\sim\; \in\{=,>,<\}$.
\item $Resets_i$ is the set of assignments of the form $c_{i_x} := 0$ or $v_{i_x} := k'$, where $c_{i_x} \in
    C_i$, $v_{i_x} \in V_i$, and $k' \in \mathbb{N}$.
\end{enumerate}

\item $Inv_i$ is the set of mode invariants mapping a mode to a
subspace of $\mathbb{R}^{|C_i|}$ indicating the possible clock
values to maintain in the mode.
\end{list1}
\end{defi}

In the following, we summarize required components of the verification model and outline our observations.
\begin{figure}
 \centering
 \includegraphics[width=0.4\columnwidth]{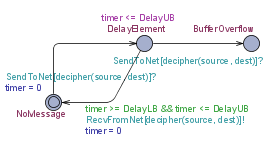}
 \caption{Timed automaton representing point-to-point transmission with capacity 1.}
 \label{fig:TA.p2p}
\end{figure}
\subsection{Network Element with Finite Capacity\label{sec.Network.Finite.Capacities}}
To model the network of the distributed system, an appropriate level of detail must be selected. In general, for a network with message delay and $n$ junction points, we have to model such a network with $n(n-1)$ automata to handle point-to-point communication. Fig.~\ref{fig:TA.p2p} is the template (defined in UPPAAL \cite{bdl:2004:uppaal}) of a timed automaton  which models the point-to-point transmission with storage capacity equal to $1$, and one overflow location. The function $\verb"decipher(source,dest)"$ is used to return the index of the channel.

\begin{obser}
For modeling of network components, only finite capacity can be reached.
Furthermore, the number of controlled locations grows exponentially as the number of allowed storage increases, because
the variable delay (due to the fault model) or the routing scheme may lead to an arbitrary ordering of arrived messages\footnote{As the
synchronizer in CTA takes no time, the timing and the ordering of messages should be modeled in the network automaton.}.
\end{obser}

\subsection{Task Element with Finite Precision\label{sec.Task.Element.Finite.Precision}}
\begin{figure}
 \centering
 \includegraphics[width=0.5\columnwidth]{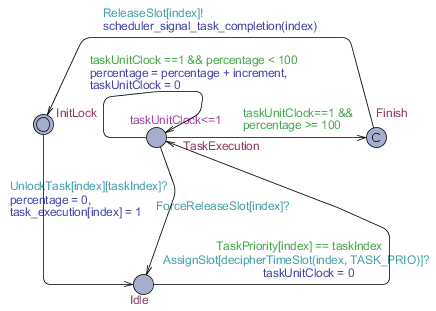}
 \caption{A timed automaton to represent task execution with context switch.}
 \label{fig:TA.task}
\end{figure}

For Giotto-like MoCs, tasks are units which perform dedicated computations. Modeling the task execution can vary based depending on whether the applied scheduling is preemptive.
Fig.~\ref{fig:TA.task} shows a timed automaton representing the task execution with potential context switches. Since context switch and preemption can occur, the
task should keep the record for the remaining time (portion) to finish the task. The variable \verb"percentage" represents the progress of execution and \verb"increment" reflects the minimal advance related to the time accuracy used during verification. The constraints regarding time accuracy imply finite precision in the model.

\begin{obser}
Modeling of tasks can only be achieved with finite precision, since with context switch, we need to record the portion
of executed tasks. This also brings issues between expressiveness and complexity;
a better accuracy regarding the timing behavior of the context switch (with finer time unit) leads to increasing complexity of the resulting model since it depends on the biggest integer used
in the system.
\end{obser}

\subsection{Job Processing Element\label{sec.Job.Processing.Element}}
The task of the job processing element is to manage the execution of tasks and to implement the inter-task communication. The construction in timed automata may vary due to the concrete application. However, for FTOS, a fixed sequence of atomic actions is defined in the software model. Each atomic action can be represented by a similar model as used for the task model described previously. The main difference is that the models of the atomic actions are linked together instead of having a closed loop in the automaton representing the job processing element. Here we omit the detailed construction process for the original model, but focus on the transformation into the according verification model. We give two motivating examples.

\begin{itemize}
    \item In order to model the effect of faults, we need to add additional edges on the original model to represent the occurrence of faults.
    \item To observe deadline violation, additional clocks that reflect the time progress since event occurrence, locations that represent the deadlines, and jumps are required to annotate the original model.
\end{itemize}

For these purposes, we define this annotation as a sequence of edit-operations over a labeled graph \cite{wang2008prs}; this facilitates the mathematical formulation how we transform between models.

\begin{defi} Define five atomic edit actions as follows\footnote{Here we merely define edit actions necessary for our propositions and algorithms; more can be defined.}.
\begin{enumerate}
\item \textbf{Clock add}: Given a clock variable $c$, $\lambda X.\emcadd(X,c)$ is an operation that adds a clock to $X$.
      Formally speaking, given  $\mathcal{A}_i = (Q_i, V_i, C_i, Sync_i,q_i,Jump_i, Inv_i)$, the result of $\emcadd(\mathcal{A}_i ,c)$ is a new timed automaton $\mathcal{A}_i' = (Q_i, V_i, C_i\cup\{c\}, Sync_i, q_i, Jump_i, Inv_i)$.
\item \textbf{Variable add}: Given a variable $v$, $\lambda X.\emvaradd(X,v)$ is an operation that adds a variable to $X$.
      Formally speaking, given $\mathcal{A}_i = (Q_i, V_i, C_i, Sync_i,q_i,Jump_i,$ $ Inv_i)$, the result of $\emvaradd(\mathcal{A}_i ,v)$ is a new timed automaton $\mathcal{A}_i' = (Q_i, V_i\cup\{v\}, C_i, Sync_i, q_i, Jump_i, Inv_i)$.
\item \textbf{Location add}
    Given a location $q$ and an invariant $inv$, where $inv$ is the conjunction of inequalities of the form $c_{i_x} \sim k$ with clock $c_{i_x}$,  $k \in \mathbb{N}$, and $\sim\; \in\{=,>,<\}$,
    $\lambda X.\emvadd(\mathcal{A}_i ,q, inv)$  is an operation that adds a location to $X$ with invariant condition $inv$.
   Formally speaking, let $\mathcal{A}_i = (Q_i, V_i, C_i, Sync_i,q_i,Jump_i, Inv_i)$, the result of $\emvadd(\mathcal{A}_i ,q, inv)$ is a new timed automaton
   $\mathcal{A}_i' = (Q_i\cup\{q\}, V_i, C_i, Sync_i, q_i, Jump_i, Inv_i\cup\{inv\})$.

\item \textbf{Jump add}:
    Given two locations $q,q'\in Q$ with guard $g$, assignment $a$, and set of synchronizers $s$, where
    \begin{enumerate}
    \item $g$ is the conjunction of inequalities of the form $c_{i_x} \sim k$ or $v_{j_x} \sim k'$, where $c_{i_x}$ is a clock, $v_{j_x}$ is a variable, $k,k' \in \mathbb{N}$, and $\sim\; \in\{=,>,<\}$.
    \item $a$ is the set of assignments of the form $c_{i_x} := 0$ or $v_{i_x} := k'$, where $c_{i_x}$ is a clock, $v_{i_x}$ is a variable, and $k' \in \mathbb{N}$.
    \end{enumerate}
    Let $\mathcal{A}_i = (Q_i, V_i, C_i, Sync_i,q_i,Jump_i, Inv_i)$, then the result of $\emjadd(\mathcal{A}_i ,q,g, a,s,q')$ is a
    new timed automaton $\mathcal{A}_i' = (Q_i, V_i, C_i, Sync_i, q_i, Jump_i\cup\{((q,g,s),(q',a))\}, Inv_i)$ by adding an arc $((q,g,s),(q',a))$ to $Jump_i$.
\item \textbf{Jump edit}:
    Given two locations $q,q'\in Q$ with guards $g,g'$, assignments $a,a'$, and sets of synchronizers $s,s'$.
    \begin{enumerate}
    \item $g, g'$ are conjunctions of inequalities of the form $c_{i_x} \sim k$ or $v_{j_x} \sim k'$, where $c_{i_x}$ is a clock, $v_{j_x}$ is a variable, $k,k' \in \mathbb{N}$, and $\sim\; \in\{=,>,<\}$.
    \item $a, a'$ are sets of assignments of the form $c_{i_x} := 0$ or $v_{i_x} := k'$, where $c_{i_x}$ is a clock, $v_{i_x}$ is a variable, and $k' \in \mathbb{N}$.
    \end{enumerate}
    Let $\mathcal{A}_i = (Q_i, V_i, C_i, Sync_i,q_i,Jump_i, Inv_i)$, then the result of $\emjedt(\mathcal{A}_i ,q,g, a,s,q',g',a',s')$ is a new timed automaton
    $\mathcal{A}_i' = (Q_i, V_i, C_i, Sync_i, q_i, Jump_i\cup\{((q,g',s'),(q',a'))\}\backslash\{((q,g,s),(q',a))\}, Inv_i)$ by changing the arc $((q,g,s),(q',a))$ to $((q,g',s'),(q',a'))$ in $Jump_i$.
\end{enumerate}
\end{defi}

Note that in our formulations, we assume due to simplification reasons that the added element is not identical to any elements in the original set, and
every newly added location or jump is well defined (e.g., to add $\mathcal{A}$ a new location with invariants using clock $c$ , $c$ should have been
defined in $\mathcal{A}$).

\begin{defi}
Let an edit sequence be $\hat{e}=e_1\circ e_2\ldots \circ e_n$, where $e_1,e_2\ldots ,e_n$ are edit actions.
Define the result of $\hat{e}$ on $\mcla$, in symbols $\mcla
e_1 \circ e_2\ldots \circ e_n$ inductively as follows.
\begin{itemize}
\item $\mcla \epsilon=\mcla$ where $\epsilon$ is the null sequence.
\item $\forall c$, $\mcla\left(\lambda X.\emcadd(X,c)\right) \circ e_2\ldots \circ e_n= \emcadd(\mcla,c)$ $\circ e_2 \ldots \circ  e_n$.
\item $\forall v$, $\mcla\left(\lambda X.\emvaradd(X,v)\right) \circ e_2\ldots \circ e_n= \emvaradd(\mcla,v) \circ e_2 \ldots \circ  e_n$.
\item $\forall q,L$, $\mcla\left(\lambda
X.\emvadd(X,q,L)\right) \circ e_2\ldots \circ  e_n = \emvadd(\mcla,q,L) \circ e_2\ldots \circ  e_n$.

\item $\forall q,q',a,s, g$, $\mcla\left(\lambda X.\emjadd(X,q,g,a,s,q')\right) \circ e_2\ldots \circ  e_n
    = \emjadd(\mcla,q,g, a,s,q') \circ e_2\ldots \circ  e_n$.
\item $\forall q,q',a,g,s,a',g',s'$, $\mcla (\lambda X.\emjedt(X,q,g,a,s,q',g',$ $a',s')) \circ e_2\ldots \circ  e_n
    = \emjedt(\mcla,q,g,a,s,q',g',a',s')$ $ \circ e_2\ldots \circ  e_n$.
\end{itemize}
\end{defi}

Starting from the textual description of the fault model, we can construct the set of \emph{deadline requirements} $\bigcup_i(q_i,q'_i,T_i)$ for the system
model $\mathcal{S}$. Intuitively this means that for all runs entering the location $q_i$, it must subsequently enter $q'_i$ within at most $T_i$ time units.
Based on above definitions, we sketch the algorithm\footnote{This editing algorithm is not general;  for requirement $(q_i,q'_i,T_i)$, $q_i$ is not reentered before entering $q'_i$ because actions in the job processing element are chained.} how to generate the verification model from the system model as follows:
\begin{small}
\procbegin
\textbf{Algorithm}: \verb"GenVerificationModelPart()"\\
\{
\hstt /* Input: Original system model $\mathcal{S}=\{\mathcal{A}_1, \ldots , \mathcal{A}_n\}$ */
\hstt /* Output: Verification model $\mathcal{S}_v$ */
\hstt \textbf{let} $\hat{e}=\epsilon$.
\hstt \textbf{forall} deadline requirements $(q_i,q'_i,T_i), q_i \in Q_i$,
\hstt /* add new clock and new variable for testing */
\hstttt$\hat{e}:=\hat{e} \circ \lambda X.\emcadd(X ,c_i)$.
\hstttt$\hat{e}:=\hat{e} \circ \lambda X.\emvaradd(X ,v_i)$.
\hstt /* $q_{dl.vio_i}$ is the location for deadline violation */
\hstttt$\hat{e}:=\hat{e} \circ \lambda X.\emvadd(X ,q_{dl.vio_i},\phi)$.
\hstttt \textbf{forall} incoming jumps $((q,g,s),(q_i,a))$ of $q_i$,
\hstttttt $\hat{e}:=\hat{e} \circ \lambda X.\emjedt(X ,q,g,s,q_i,a, g, a',s)$,
\hstttttttt where $a'= a \wedge (c_i := 0)\wedge(v_i:=0)$.
\hstttt \textbf{endfor}
\hstttt \textbf{forall} incoming jumps $((q,g,s),(q'_i,a))$ of $q'_i$,
\hstttttt $\hat{e}:=\hat{e} \circ \lambda X.\emjedt(X ,q,g,s,q'_i,a, g, a',s)$,
\hstttttttt where $a' = a \cup\{(v_i:=1)\}$.
\hstttt \textbf{endfor}
\hstttt \textbf{forall} reachable locations $q$ from $q_i$,
\hstttttt $\hat{e}:=\hat{e} \circ \lambda X.\emjadd(X ,q,g,\phi,\phi, q_{dl.vio_i})$.
\hstttttttt where $g$ is defined as $(v_i = 0) \wedge (c_i>T_i)$.
\hstttt \textbf{endfor}
\hstt \textbf{endfor}
\hstt \textbf{return} $\mathcal{S}_v := \mathcal{S}\hat{e}$. /* apply changes in $\hat{e}$ */
\\\}
\procend
\end{small}

For the property of deterministic assumption mentioned in section~\ref{subsec.Verification.Goals}, similar algorithms can be applied to annotate clocks, locations, and jumps; the problem for checking deterministic assumption in FTOS turns to be a reachability problem in timed automata.

\subsection{Dispatcher\label{sec.Dispatcher}}
With respect to the operating system, we have to model the dispatcher explicitly.
The modeled dispatcher merely captures the scheme for the execution of threads; deadline violation, fault-tolerance or error handling is modeled in the job processing element. Therefore, it can be used in arbitrary settings and not only in FTOS.
Due to different scheduling algorithms, the model of the dispatcher differs dramatically regarding actual verifiability.
For our analysis, we use priority based dispatchers modeling either FIFO or round-robin techniques.
Nevertheless, as context switch of tasks/threads occurs, we have the following observation.

\begin{obser}
Using a round-robin dispatcher leads to exponential increase of possible behaviors compared to a FIFO-based dispatcher with the number of parallel tasks, if no assumptions on the task behavior can be made.
\end{obser}

In summary, this section gave insight in the main components of the verification model and their construction. Besides the job processing element, all components and related observations can be directly applied for arbitrary real-time systems. For the job processing element, we described a generic way to use annotations to construct a model to use for verifying the absence of deadline violations. In the next section, we point out how aperiodic behavior introduced by faults or events can be considered.
\section{Invocation of Faults and Aperiodic Events\label{Sec.Invocation.Faults.Events}}

\begin{figure}
 \centering
 \includegraphics[width=0.15\columnwidth]{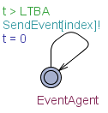}
 \caption{A sample timed automaton representing the event agent.}
 \label{fig:TA.event.agent}
\end{figure}

To perform verification, modeling the arrival of faults or aperiodic events is necessary to establish a closed model, and in this section we consider its effect.
In FTOS, the probability of faults is implicitly reflected by the concept called \emph{least time between faults} (LTBF). In our analysis, the invocation of aperiodic tasks can be done similarly - the least time between occurrences
of events for aperiodic tasks is defined as \emph{least time between arrivals} (LTBA). With LTBA or LTBF,
we can augment the original model with a timed automaton producing the event (called \emph{event agent}) similar to fig.~\ref{fig:TA.event.agent}.
However, since LTBF (or LTBA) is an integer which might be relatively large, and the complexity of verification in timed systems is related to this
integer\footnote{The reachability problem for timed automata is PSPACE-complete, i.e., the complexity is exponential to (1) the number of clocks and (2) the
maximum integer used in the system. Concerning (2), if the maximum number changes from $10$ to $100$, intuitively the execution time can increase by the factor of $k^{90}$, where $k>1$.},
the use of LTBF (or LTBA) may hinder the practicability of model checking.
Thus we propose some methods to effectively reduce the value of LTBF (or LTBA) with equivalence criterion.
For simplicity reasons, the following theorems are all discussed using event-triggered aperiodic functions with LTBA without loss of generality.

\begin{prop}\label{simple}
Let system $S$ have one FTOS function with periodic deadline $T$ and one event-triggered aperiodic function. 
\begin{list1}
\item W.L.O.G., let $A_{event}=(\{q\},\emptyset,t,\{!event\},q,\{((q,(t>LTBA_{S}),\{!event\}),(q,(t:= 0))),\phi\})$ be the timed automaton of the event agent,
where $LTBA_{S}=T_S$ be the least time between two consecutive aperiodic events.
\end{list1}
Let $T_p$ be the maximal time interval for the system to finish processing the event (called deadline interval from now on)\footnote{Let $t_{arrival}$ be the time for the event arrival. If the system can not finish processing this event within time $t_{arrival}+T_p$, then the system violates the deadline.}. If $T_{S} > T_{p} + T$, then consider another system $S'$, where $S' = \emjedt(S ,q,(t>T_S),q,(t:= 0), \{!event\},(t>T_{p} + T), (t:= 0), \{!event\})$, i.e., the
only difference is to change $LTBA_{S}$ from $T_S$ to $T_{p} + T$. Then both systems are equivalent regarding their behavior concerning deadline violation.
That is, for $S$ and $S'$, either they both satisfy the deadline, or they both miss the deadline.
\end{prop}

An intuitive argument for the bound $T_p+T$ can be derived using fig.~\ref{fig:Proof.Explanation}. Tasks and events influence the execution of each other. The execution of an arriving event is influenced by the currently running task. During the deadline interval of the event, this and all preceding tasks are influenced as well. The chain of influence can only be stopped if the execution is decoupled. Since preceding tasks are decoupled by definition, two events with a minimal bound of $T_p+T$ can not influence the execution of the same task. In fig.~\ref{fig:Proof.Explanation}, we call a time point $\hat{t}$ \emph{decoupling point} if two consecutive tasks immediately before and after $\hat{t}$ are not mutually influenced due to the occurrence of an event.

\begin{IEEEproof}
We consider four possible cases in $S'$:
\begin{enumerate}
    \item Consider the case where in $S'$, it is proven that no deadline is violated. When the verification engine proofs that the deadline is never
            violated with $LTBA_{S'}=T_{S'}$ in $S'$, the deadline of the FTOS function in $S$ will never be violated because $T_{S} > T_{S'}$; the verification engine has already considered  all cases in $S$.
    \item Consider the case where in $S'$, the counter-example indicates that the $i$-th aperiodic task violates the deadline. We further split the discussion in subjects whether it is the first time for $S'$ to process the event. Our goal is to construct a counter-example for deadline violation in $S$ from the counter-example in $S'$.
\begin{figure}
 \centering
 \includegraphics[width=0.5\columnwidth]{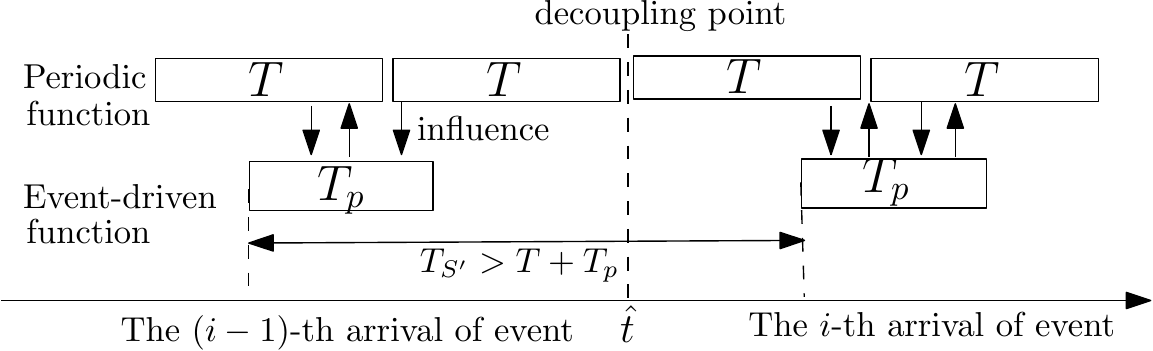}
 \caption{Illustrations for proofs of Proposition~\ref{simple}.}
 \label{fig:Proof.Explanation}

\end{figure}
        \begin{enumerate}
            \item If $i=1$, i.e., it is the first time for $S'$ to execute the aperiodic task, then this deadline violation can also occur in $S$, since no constraints are made for the first
                occurrence of events in $S$ or $S'$.
            \item If $i \neq 1$, consider the $(i-1)$-th aperiodic execution which does not violate the deadline. Let the time for the coming of event $(i-1)$-th be $t$, and let the interval between the $(i-1)$-th
                  and the $i$-th event be $T'$.  The system should finish the $(i-1)$-th processing before time $t+T_p$. Since $T'>LTBA_{S'}=T_p + T$, from time $t+T_p$ to $t+T'$, FTOS function should finish one of its execution and proceed a new one. Let the time for the start of that cycle be $\hat{t}$. If we change the counter-example time trace such that no event has happened before $\hat{t}$, we still get a counter-example trace in $S'$. This new counter-example trace is also a counter-example trace in $S$.
        \end{enumerate}
    \item Consider the case where in $S'$, the counter-example indicates that the FTOS function violates the deadline. Let the time which violates the deadline be $t$ (note that $t$ is the multiple of $T$). Let the occurrence of the nearest event be $t'$ (if there exists no such event, then both $S$ and $S'$ can deadlock).
          \begin{enumerate}
            \item  If $t - t' \geq T_p + T$, then the event is processed before time $t-T$, the starting of the period which violates the deadline. In this way, the system violates the deadline with only the existence of FTOS function, thus in $S$, the deadline will also be violated.
            \item  If $t - t' < T_p + T$, we consider whether the event is the first one being processed.
            \begin{enumerate}
            \item If yes, then the counter-example in $S'$ is also a counter-example in $S$.
            \item If not, then consider the time where the previous event occurs, and let the time be $t''$. Since $t''-t'> T+ T_p$, we can find a decoupling point $\hat{t}$, where $t'' + T_p\leq \hat{t} \leq t'$, where at $\hat{t}$ it starts a new period.
                In this way, we can perform the same technique stated in (2-b) before $\hat{t}$.
            \end{enumerate}
          \end{enumerate}
    \item Consider the case where in $S'$, the counter-example indicates that both the $i$-th aperiodic task and the FTOS function violate the deadline.
          Let the time which violates the deadline be $t$ (note that $t$ is the multiple of $T$), then the event occurs in time $t-T_p$. By an argumentation similar to point 3-b, a counter-example trace in $S$ can be established.
\end{enumerate}

\end{IEEEproof}

Remark: (1) Proposition~\ref{simple} formulates the insight that previous events occurred long before can not influence the current processing and scheduling, and therefore, are not the root cause of deadline violation. In other words, we could also construct a counter example with a single event as root cause.
(2) The introduction of faults can be viewed analogously. For faults, in FTOS (or similar fault-tolerant systems) the value of $T_p + T$ is much smaller then $T_S$ (LTBF), and this brings
significant advantages for construction of a model with smaller state space.

Proposition~\ref{simple} can only be used for very simple systems with only one task and one event. In the following, we will generalize the result to systems consisting of one periodic function with period $T$ and several aperiodic functions.

\begin{prop}\label{more.events}
 Let $S$ be a system  with $n$ aperiodic functions. Each function with index $i$, where $i= 1\ldots n$, is associated with a pair
 $(LTBA_i, T_{p_i})\in \mathbb{N}\times\mathbb{N}$ describing the LTBA and deadline interval.
 Consider another system $S'$, where the only difference is to perform the following change: if for all $i= 1\ldots n$,
$LTBA_i > T(\sum_{i= 1 \ldots n}\lceil\frac{ T_{p_i}}{T}\rceil) + T$, then we change $LTBA_i$ to $T(\sum_{i= 1 \ldots
n}\lceil\frac{ T_{p_i}}{T}\rceil) + T$. Both systems $S$ and $S'$
are equivalent regarding their behavior concerning deadline
violation.
\end{prop}

\begin{IEEEproof} We consider the following cases.
\begin{enumerate}
    \item If $S'$ does not violate the deadline, then so does $S$.
    \item Consider the case where in $S'$, the counter-example indicates that the $i$-th aperiodic task of type
          $j$ violates the deadline.
    \begin{enumerate}
          \item If $i \neq 1$, let the time for the $i$-th and $(i-1)$-th arrival of type-$j$ events be $t$
          and $t'$. Our goal is to find the decoupling point $\hat{t}$ such that we can overlook
          all previously happened events.

          Since $t-t'>T(\sum_{i= 1 \ldots n}\lceil\frac{ T_{p_i}}{T}\rceil) + T$,
          then within $[t',t]$ the periodic function is executed at least $\alpha=(\sum_{i= 1 \ldots n}\lceil\frac{ T_{p_i}}{T}\rceil + 1)-1$
          times. Consider the worst case where it is only executed
          $\alpha$ times. Within $[t',t]$, there are $\alpha+1$ potential decoupling points $t_{0},\ldots, t_{\alpha+1}$, where $\forall k=1\ldots \alpha+1$, $t_{k}-t_{k-1}=T$.

          Due to the sparsity of events, each type of event arrives at most once within $[t',t]$. For each type $m$, the according event with deadline interval $T_{p_m}$ will overlap in worst case at most $\lceil\frac{ T_{p_i}}{T}\rceil$ of these potential decoupling points.
          Thus the total number of overlapped points is at most $\sum_{i= 1 \ldots n}\lceil\frac{ T_{p_i}}{T}\rceil = \alpha$,
          which is less than the number of points among $\{t_{0},\ldots, t_{\alpha+1}\}$. Therefore,
          there exists at least one point $t_{g}\in \{t_{0},\ldots, t_{\alpha+1}\}$ such that it is not overlapped
          by any deadline interval. Thus we can set the decoupling point $\hat{t}$ as $t_{g}$. As a result, we can construct an equivalent counter-example where no event has happened before $\hat{t}$. This new counter-example trace is also a counter-example
            trace in $S$.
          \item If $i = 1$, let the time for the $i$-th  arrival of type-$j$ events be $t$.
            \begin{enumerate}
                \item If for all type of events, the according events occurred at most once before $t$, then the counter-example
                is also a counter-example in $S$.
                \item If there exists some type of events occurred more than once: let $c_m$ for type $m$  be
                the total number of events occurred in the counter-example and $t_{c_m}$ be the latest event
                arrival time. Choose $m'$ such that $c_m>1$ and $\forall m=1\ldots n, m\neq j$, $t_{c_{m'}}>t_{c_m}$.
                Then we can find the decoupling point between the $(c_{m'}-1)$-th and $(c_{m'})$-th arrival of
                event with type $m'$, similar to the argument in case 2-a.
            \end{enumerate}
    \end{enumerate}
    \item Consider cases where the deadlock happens in the FTOS function.
    \begin{enumerate}
        \item If in the counter-example no event has occurred, then both $S$ and $S'$ can deadlock.
        \item Otherwise, first we try to pick an event based on arguments in 2-b-ii. If possible, then the decoupling point can be found, and the counter-example for $S$ can be established. If selection based on 2-b-ii is not possible, it follows the statement of 2-b-i that the counter example for $S'$ is also a counter-example in $S$.
    \end{enumerate}

\end{enumerate}
\end{IEEEproof}

Lastly, we discuss the most general case.

\begin{prop}\label{more.tasks}
Let system $S$ have $m$ periodic FTOS/Giotto functions with periodic deadline $T_{f_1}, T_{f_2},\ldots T_{f_m}$, where $i= 1\ldots m$, and
$n$ aperiodic functions. Each function with index $j$, where $j= 1\ldots n$, is associated with a pair
 $(LTBA_j, T_{p_j})\in \mathbb{N}\times\mathbb{N}$ describing the LTBA and deadline interval.
 Consider another system $S'$, where the only difference is to perform the following change: if for all $i= 1\ldots n$,
$LTBA_i > T'(\sum_{i= 1 \ldots n}\lceil\frac{ T_{p_i}}{T'}\rceil) + T'$, where $T'$ is the least common multiple of
$T_{f_1}, T_{f_2},\ldots T_{f_m}$, then we change $LTBA_i$ to $T'(\sum_{i= 1 \ldots
n}\lceil\frac{ T_{p_i}}{T'}\rceil) + T'$. Both systems $S$ and $S'$
are equivalent regarding their behavior concerning deadline
violation.
\end{prop}

\begin{IEEEproof}The main difference to the previous case is that periodic functions with different tasks might influence each other. Potential decoupling points occur only at points in time, where all tasks start together. The proof idea is to view multiple periodic functions as a whole by taking the least common multiple. Here we omit the detailed proof.
\end{IEEEproof}


\section{Implementation\label{sec.Implementation}}

For implementation, we extend the functionality of FTOS-Verify to test the applicability. The verification model is constructed in a format acceptable by UPPAAL~\cite{bdl:2004:uppaal}.
Note that templates in UPPAAL are not completely suitable for our usage, since they only represent a fixed behavior with configurable
parameters. Therefore, algorithms to automatically generate timed automata based on FTOS models are needed. We have implemented our automated M2M transformation tool using openArchitectureWare\footnote{http://www.openarchitectureware.org}under the Eclipse modeling framework\footnote{http://www.eclipse.org/modeling/emf/}.

As use case, we apply the verification in the context of our balanced-rod example\footnote{For configurations, see http://www6.in.tum.de for details.}, where the control functions are replicated on three redundant machines to guarantee fault-tolerance. All components mentioned previously are generated by our automatic conversion technique; the resulting UPPAAL system has $25$ communicating timed automata. As timing information for the different components, we use currently user-specified assumptions. An integration of WCET-analyzers is foreseen. One desired property specified is the guarantee for the absence of deadline violation, which turns to be the reachability property in UPPAAL.

The overall execution time varies from $1$ to $25$ minutes depending on the accuracy of the verification model on a Intel 2.33 GHz machine using a FIFO-based priority-driven scheduler. The  memory consumption can reach up to $850$Mb. The verification of using a Round-Robin based scheduler showed to be too memory comsuming.

\section{Related Work\label{sec.Related.Work}}
We mention related work, but constrain ourselves in works regarding the analysis of Giotto-like systems; for techniques applying formal verification
in real-time analysis, we refer readers to the survey paper by Wang \cite{wang2004fvt}.
In Giotto, the Giotto-Compiler will perform hardware mapping and apply analysis techniques to check schedulability.
Many design tools with Giotto-like MoCs apply similar approaches, for example, TDL~\cite{simmons:1998:tdl} or HTL~\cite{ghosal:2006:htl}, but
analysis techniques are not explicitly mentioned.
One interesting work comes from COMDES-II project~\cite{ke2008vci}, which is also based on the concept of logical execution time; here, researchers apply model transformation from system models to verification models.
Nevertheless, as we focus on fault-tolerant systems, our work differs from the above works with the following facts.
First, we encounter a harder problem; by applying software fault-tolerance, modeling the communication between multiple deployed
units is required, and this is not required by other Giotto-like MoCs.
For those MoCs, scheduling analysis developed in real-time community could be enough without the use of model checking.
Furthermore, by proposing the similarity between aperiodic events and fault occurrences, our theoretical criteria is powerful to
reduce dramatically the complexity of the model (not the verification algorithm). This is based on our understanding regarding constituents
for the complexity of timed verification.

\section{Conclusion\label{sec.Conclusion}}

In this paper, we discussed the issue of constructing a model to verify timing assumptions in the context of FTOS using timed automata. However, due to our general approach, the results can be applied to arbitrary distributed real-time systems.

Our contribution can be summarized as follows.
\begin{enumerate}
    \item We give observations concerning modeling of general distributed real-time systems using timed automata and formulate our verification model construction process.
    \item With the context of systems consisting of periodic and aperiodic tasks, we give theoretical criteria how to reduce the size of the verification model, which is particularly useful for our approach. The change of the maximum integer used in the system decreases the required time for verification with exponential scale.
    \item A prototype software for the conversion process is constructed with preliminary experiments.
\end{enumerate}

Our work is currently based on user-specified assumptions regarding the timing of involved components. The next step will be the integration of WCET analyzing tools to have a faithful verification result.

Furthermore, we are investigating on approaches to separate the verification problem for control functions executed in parallel to make our approach applicable also for large-scale applications.


\bibliographystyle{plain}

\end{document}